\documentclass[ras]{AGUTeX}

\usepackage[dvips]{graphicx}

\usepackage{rotating}
\setkeys{Gin}{draft=false}
\authorrunninghead{SANE ET AL.}

\titlerunninghead{DATAFLOW MODELS FOR RADIO ASTRONOMY DSP}

\authoraddr{S. S. Bhattacharyya,
Department of Electrical and Computer Engineering, and
Institute for Advanced Computer Studies,
University of Maryland, College Park, MD, 20742, USA.
(ssb@umd.edu)}

\authoraddr{J. Ford, 
National Radio Astronomy Observatory, 
Green Bank, WV, 24944, USA.
(jford@nrao.edu)}

\authoraddr{A. I. Harris,
Department of Astronomy,
University of Maryland, College Park, MD, 20742, USA.
(harris@astro.umd.edu)}

\authoraddr{N. Sane,
Department of Physics, and
Center for Solar-Terrestrial Research,
New Jersey Institute of Technology, Newark, NJ, 07102, USA.
(nimish.sane@njit.edu)}

\begin{document}

%

\title{Prototyping scalable digital signal processing systems for radio
  astronomy using dataflow models}
%

%
%



\authors{N. Sane, \altaffilmark{1}
J. Ford, \altaffilmark{2} A. I. Harris, \altaffilmark{3}
and S. S. Bhattacharyya\altaffilmark{4}}

\altaffiltext{1}{N. Sane was with the Department of Electrical and
  Computer Engineering, and Institute for Advanced Computer Studies at
  University of Maryland, College Park, MD, USA. He is now with the
  Department of Physics, and Center for Solar-Terrestrial Research at
  New Jersey Institute of Technology, Newark, NJ, USA.}

\altaffiltext{2}{National Radio Astronomy Observatory, 
Green Bank, West Virginia, USA.}

\altaffiltext{3}{Department of Astronomy,
University of Maryland, College Park, Maryland, USA.}

\altaffiltext{4}{Department of Electrical and Computer Engineering, and
Institute for Advanced Computer Studies,
University of Maryland, College Park, Maryland, USA.}

%
%


\begin{abstract}
There is a growing trend toward using high-level tools for design and implementation of radio astronomy digital signal processing (DSP) systems. Such tools, for example, those from the Collaboration for Astronomy Signal Processing and Electronics Research (CASPER), are usually platform-specific, and lack high-level, platform-independent, portable, scalable application specifications. This limits the designer's ability to experiment with designs at a high-level of abstraction and early in the development cycle.

We address some of these issues using a model-based design approach employing dataflow models. We demonstrate this approach by applying it to the design of a tunable digital downconverter (TDD) used for narrow-bandwidth spectroscopy. Our design is targeted toward an FPGA platform, called the {\em Interconnect Break-out Board} ({\em IBOB}), that is available from the CASPER. We use the term TDD to refer to a digital downconverter for which the decmation factor and center frequency can be reconfigured without the need for regenerating the hardware code. Such a design is currently not available in the CASPER DSP library.

The work presented in this paper focuses on two aspects. Firstly, we introduce and demonstrate a dataflow-based design approach using the {\em dataflow interchange format} ({\em DIF}) tool for high-level application specification, and we integrate this approach with the CASPER tool flow. Secondly, we explore the trade-off between the flexibility of TDD designs and the low hardware cost of fixed-configuration digital downconverter (FDD) designs that use the available CASPER DSP library. We further explore this trade-off in the context of a two-stage downconversion scheme employing a combination of TDD or FDD designs.
\end{abstract}

%
%

%

\begin{article}

%
%

\section{Introduction}
\label{sec:intro}
Key challenges in designing digital signal processing (DSP) systems
employed in the field of radio astronomy arise from the need to
process very large amounts of data at very high rates arriving from
one or more telescopes. It is also desirable to have scalable and
reconfigurable designs for shorter development cycles and faster
deployment. Moreover, these designs should be portable to different
platforms to keep up with advances in new hardware
technologies. However, conventional design methodologies for signal
processing systems in the field of radio astronomy focus on custom
designs that are platform-specific. Such designs, by virtue of being
platform-specific, are highly specialized, and thus difficult to
retarget. Traditional design approaches also lack high-level
platform-independent application specifications that can be
experimented with, and later ported to and optimized for various
target platforms. This limits the scalability, reconfigurability,
portability, and evolvability across varying requirements and
platforms of such DSP systems.

A model based approach for design and implementation of a DSP system
can effectively exploit the semantics of the underlying models of
computation. This facilitates precise estimation and optimization of
system performance and resource requirements (e.g.,
see~\citep{bhat2010x1}). Though approaches for scalable and
reconfigurable design based on modular field programmable gate array
(FPGA) hardware and software libraries have been developed (e.g., see
~\citep{pars2005x1, pars2006x1, szom2011x1, nallurl, lyrturl}), they
do not provide forms of high-level abstraction that are linked to
formal models of computation.

We propose an approach using DSP-oriented dataflow models of
computation to address some of these
issues~\citep{lee1987x2}. Dataflow modeling is extensively used in
developing embedded systems for signal processing and communication
applications, and electronic design automation~\citep{bhat2010x1}. Our
design methodology involves specifying the application in the dataflow
interchange format (DIF)~\citep{hsu2005x2} using an appropriate
dataflow model. This application specification is transformed into an
intermediate, graphical representation, which can be further processed
using graph transformations.

The DIF tool allows designers to verify the functional correctness of
the application, estimate resource requirements, and experiment with
various dataflow graph transformations, which help to analyze or
optimize the design in terms of specific objectives.  The DIF-based
dataflow specification is then used as a reference while developing a
platform-specific implementation. We show how formal understanding of
the dataflow behavior from the software prototype allows more
efficient prototyping and experimentation at a much earlier stage in
the design cycle compared to conventional design approaches.

We demonstrate our approach using the design of a tunable digital
downconverter (TDD) that allows fine-grain spectroscopy on narrow-band
signals. A primary motivation behind a TDD design is to support
changes to the targeted downsampling ratio without requiring
regeneration of the corresponding hardware code. Development of such a
TDD is a significant contribution of this work. We compare our TDD
with the fixed-configuration digital downconverter (FDD) designs that
use the current DSP library from the Collaboration for Astronomy
Signal Processing and Electronics Research (CASPER)
(see~\citep{casperurl}). We explore trade-offs between the flexibility
offered by TDD designs and their hardware cost. A TDD is particularly
useful since our target FPGA hardware platform --- {\em interconnect
break-out board} ({\em IBOB})~\citep{pars2006x1} --- does not have the
feature of storing more than one configurations (also referred to as
``personalities'') and dynamically loading one of them, unlike some of
the CASPER hardware platforms of a later generation. A single
reconfigurable TDD design also simplifies code management when
compared to multiple static designs.

We must emphasize that this paper describes a dataflow-based design
flow for prototyping radio astronomy DSP systems. This approach is not
restricted to any particular tool or hardware platform. We intend to
demonstrate it by developing a high-level DIF prototype that uses
dataflow formalisms and generating a hardware implementation using
CASPER tools from this DIF prototype. The proposed approach is not
intended to replace the CASPER tools. It offers enhancements to the
existing CASPER design flow. However, this does not restrict its use
to only the CASPER tools.

The organization of the rest of this paper is as
follows. Section~\ref{sec:tdd-app} describes a TDD
application. Section~\ref{sec:df} describes dataflow modeling in
detail, along with some of the relevant forms of dataflow ({\em
dataflow models}) that are employed in practice. A reader who is
familiar with dataflow formalisms may skip this
section. Section~\ref{sec:dif} provides information about the DIF
tool, while Section~\ref{sec:work} highlights some of the relevant
prior work. Section~\ref{sec:design-flow} explains how a DIF prototype
can be used to develop a hardware
implementation. Section~\ref{sec:conclusion} provides a summary and
our conclusions.

\section{Tunable Digital Downconverter}
\label{sec:tdd-app}
\begin{figure}[hb]
\centering
\noindent\includegraphics[scale=0.6]{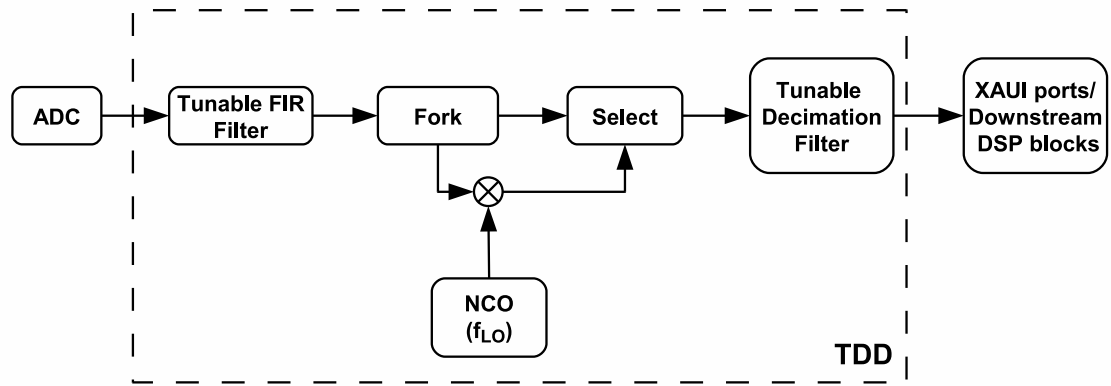}
\caption{Block diagram of a tunable digital downconverter.}
\label{fig:tdd}
\end{figure}

\begin{figure}[hb]
\begin{minipage}[hb]{1.0\linewidth}
\centering
\noindent\includegraphics[scale=0.5]{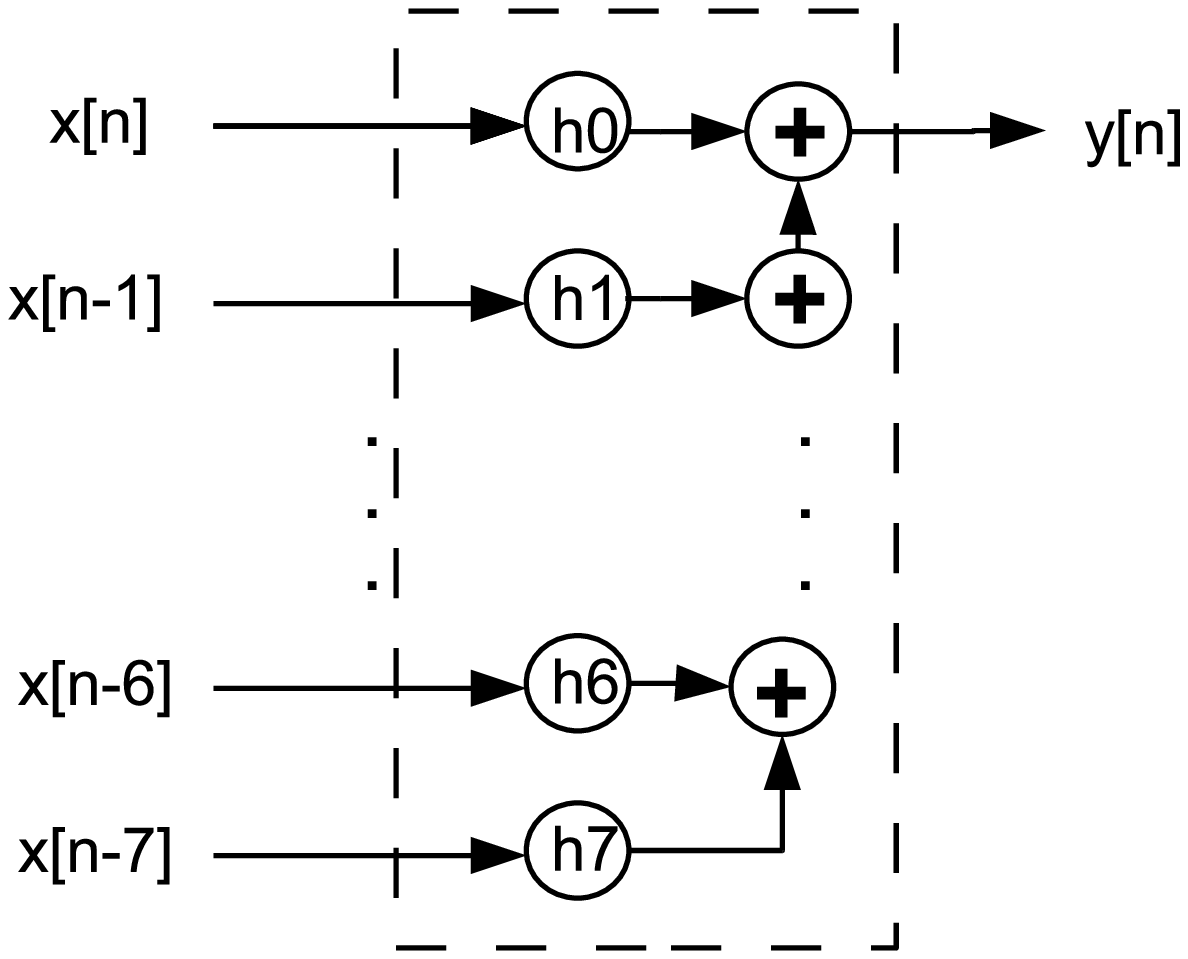}
\centering \centerline{(a)}
\end{minipage}
\begin{minipage}[hb]{1.0\linewidth}
\centering
\noindent\includegraphics[scale=0.5]{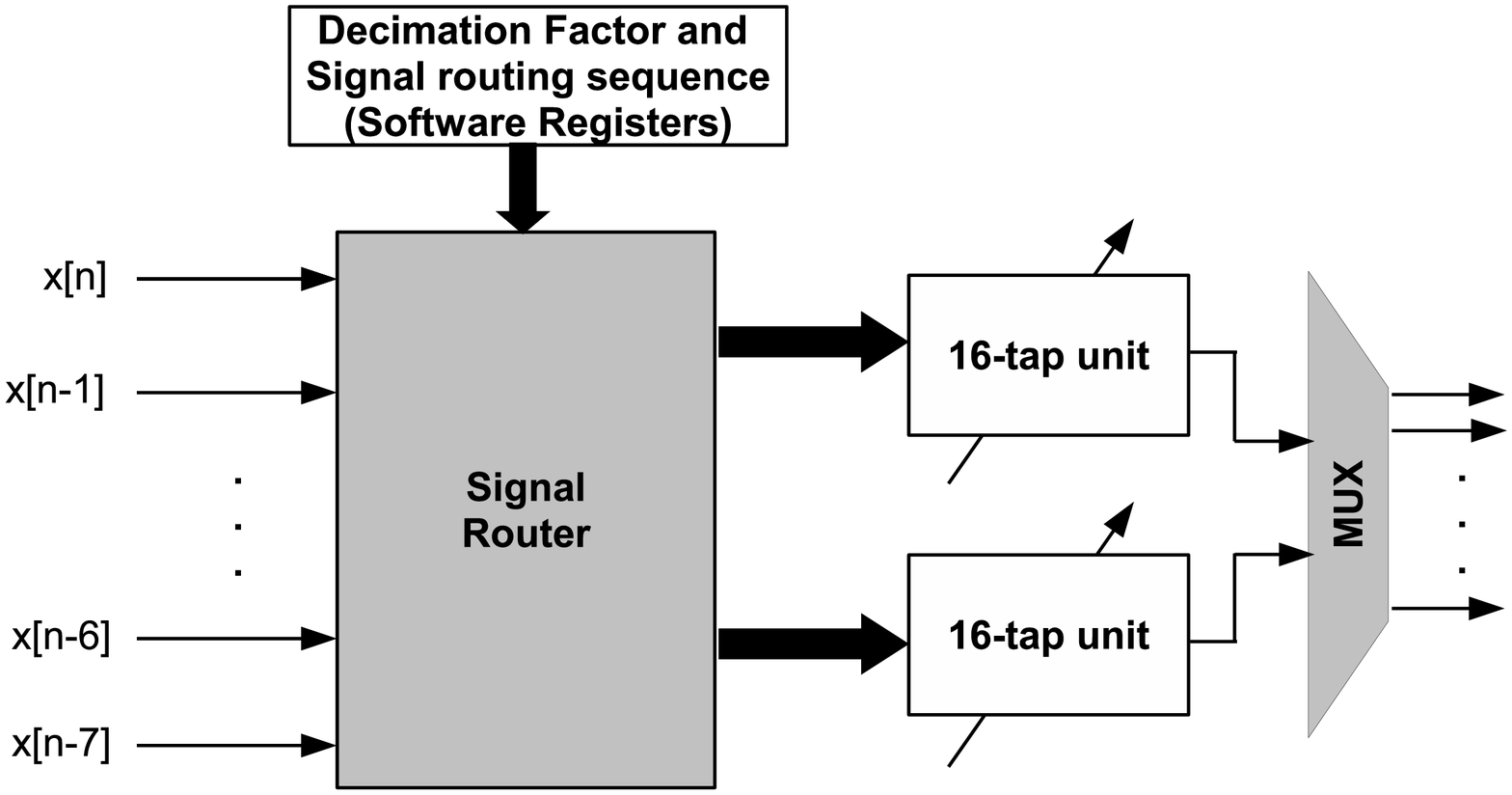}
\centering \centerline{(b)}
\end{minipage}
\caption{Schematic of (a) fixed-configuration decimation filter (FDF)
  in the CASPER library, and (b) tunable decimation filter (TDF) that
  is part of a TDD. The FDF achieves downconversion of $8$ by having
  $8$ parallel inputs $x[n], x[n-1], \ldots, x[n-7]$. Here, $h0, h1,
  \ldots, h7$ denote the filter coefficients, and $y[n]$ denotes the
  output. For TDF, 16-tap units are similar to the structure inside
  the dotted box shown in (a) with tunable filter taps. The TDF block
  has $8$ inputs as well as $8$ outputs.}
\label{fig:schem}
\end{figure}

In the DSP literature, the terms downsampling and decimation are often
used interchangeably. In this paper, a {\em decimator} refers to a
block that simply decimates or downsamples the input signal without
any other processing (e.g., see Fig.~\ref{fig:back-csdf}(a) and
(b)). The ratio of the sampling rate at the input of a decimator to
that at its output is referred to as its {\em decimation factor}. A
decimator is generally preceded by an anti-aliasing
filter~\citep{vaid1990x1}. In this paper, we refer to such a combined
structure, consisting of a filter and decimator, as a {\em decimation
filter} (e.g., see Fig.~\ref{fig:back-pcsdf}(a) and (b)). In a
polyphase implementation of a decimation filter, such as the one we
use in our implementation, this structure is implemented as a single
computing block~\citep{vaid1990x1}. We refer to the system or
application that employs a decimator or decimation filter, possibly
with other blocks such as mixers and filters, as a digital
downconverter, and in particular, a FDD or TDD (e.g., see
Fig.~\ref{fig:tdd} and Fig.~\ref{fig:schem}). The decimation factor of
a decimation filter, TDD, or FDD refers to that of the decimator in
it.
 
Fig.~\ref{fig:tdd} shows a block diagram of a TDD application. An
$8$-bit analog-to-digital converter (ADC) receives a baseband input IF
signal of bandwidth $800$\,MHz and samples it at the sampling rate of
$1.6$\,giga-samples/second (GS/s).  The internal design of the ADC
block is such that $8$ consecutive time samples, where each sample is
an $8$-bit fixed point number, are output on the eight $8$-bit buses
at the same clock pulse. This results in $200$\,mega-samples/second
(MS/s) on each of the outputs of the ADC block. Correspondingly, all
the downstream blocks also have $8$ input and output ports. Thus,
there are $8$ connections between any two blocks shown in
Fig.~\ref{fig:tdd} that are directly connected. We have not shown all
$8$ connections in detail for the sake of clarity and simplicity.

\begin{figure}[hb]
\centering
\noindent\includegraphics[scale=0.6]{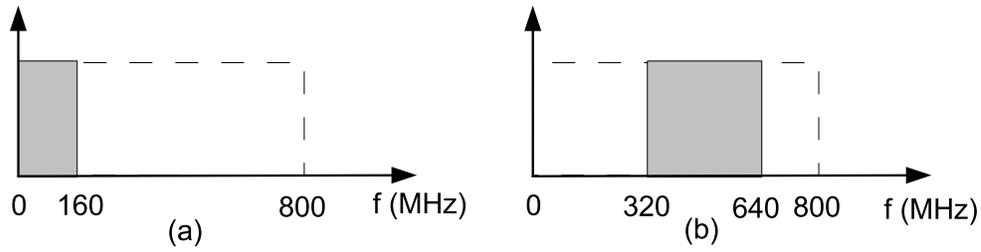}
\caption{Two of the possible configurations of a TDD: (a) $B_w = 160$
  MHz, $C_f = 80$ MHz (b) $B_w = 320$ MHz, $C_f = 480$ MHz. The
  colored area shows the extracted frequency band.}
\label{fig:tdd-band}
\end{figure}

The TDD subsystem, identified by the dotted box in Fig.~\ref{fig:tdd},
extracts a subband of the input signal with a user-specified center
frequency ($C_f$) and bandwidth ($B_w$), downconverts it to a
baseband, and then downsamples it to the Nyquist rate. For example,
Fig.~\ref{fig:tdd-band} shows two of the possible configurations of
$B_w$ and $C_f$ and the corresponding frequency bands that are
extracted. The output of the TDD can be used by the downstream DSP
blocks. For example, a possible scheme can have a TDD implementation
on the IBOB. The downstream DSP blocks may include functions such as
polyphase filtering and fast Fourier transform. These blocks can be
implemented on a different hardware. This is possible using a
communication link between two hardware boards that behaves as a FIFO
buffer. An Ethernet link using $10$x auxiliary user interface (XAUI)
ports available on the IBOB is an example of such a link.

During narrow-band observations, the Nyquist sampled output of the TDD
will be analyzed with an existing spectrometer. The same number of
spectral channels will thus provide proportionately greater spectral
resolution as compared to analyzing the entire input bandwidth. Our
TDD design supports integer decimation factors between $5$ and
$12$. The choice of these values stems purely from the initial
specification of the Green Bank Ultimate Pulsar Processing Instrument
(GUPPI)~\citep{ford2010x1}. This should be considered simply as a
demonstrative implementation. The approach presented in this paper
does not restrict the design in any way from having different
specifications. The valid values of $C_f$ corresponding to the
selected $B_w$ can vary so as to span the entire $800$\,MHz IF input.

As shown in Fig.~\ref{fig:tdd}, the TDD includes a tunable finite
impulse response (FIR) filter. If the desired output is a baseband
signal, then the FIR filter simply acts as a low-pass filter. Also, in
this case, the fork (which can be viewed as a dataflow version of a
signal splitting block) and select (which is similar to a multiplexer)
blocks are configured to route the output of the FIR filter directly
to the tunable decimation filter (TDF), bypassing the mixer.

If the desired output is not a baseband signal, the FIR filter acts as
a bandpass filter (BPF). The cut-off frequencies for this BPF are set
using the specified parameter configuration ($B_w$ and $C_f$). In this
case, the output of the BPF is fed to a real mixer, which translates
it into a baseband signal. The local oscillator, with a frequency
$f_\mathrm{LO}$, is implemented as a numerically controlled oscillator
(NCO). The frequency, $f_\mathrm{LO}$, is dependent on the value of
$C_f$ and $B_w$. The output of the mixer is then fed to the TDF, which
downsamples its input depending upon the specified $B_w$ or decimation
factor. We have used this scheme in order to have a real-valued TDF
output.

Such a TDD, which was originally designed for the GUPPI at the
National Radio Astronomy Observatory (NRAO), Green Bank, finds its use
in the spectrometers currently under development for the Green Bank
telescope (GBT) and 20m telescope at the NRAO, Green Bank.

\section{Background}
\label{sec:background}
\subsection{Dataflow Modeling}
\label{sec:df}
Dataflow modeling involves representing an application using a
directed graph $G (V,E)$, where $V$ is a set of vertices (nodes) and
$E$ is a set of edges. Each vertex $u \in V$ in a dataflow graph is
called an {\em actor}, and represents a specific computational block,
while each directed edge $(u,v) \in E$ represents a first-in-first-out
(FIFO) buffer that provides a communication link between the {\em
  source} actor $u$ and the {\em sink} actor $v$. A dataflow graph
edge $e$ can also have a non-negative integer {\em delay},
$\ensuremath{\mathrm{del}(e)}$, associated with it, which represents
the number of initial data values ({\em tokens}) present in the
associated buffer. Dataflow graphs operate based on {\em data-driven
  execution}, where an actor can be executed ({\em fired}) whenever it
has sufficient amounts of data (numbers of ``samples'' or data
``tokens'') available on all of its inputs. Typically, in DSP-oriented
data flow design environments, the execution of a dataflow graph can
be thought of as that of a ``globally asynchronous locally
synchronous'' (GALS) system~\citep{suha2008x1, shen2009x1}.

During each firing, an actor consumes a certain number of tokens from
each input and produces a certain number of tokens on each
output. When these numbers are constant (over all firings), we refer
to the actor as a {\em synchronous dataflow} ({\em SDF})
actor~\citep{lee1987x2}.  For an SDF actor, the numbers of tokens
consumed and produced in each actor execution are referred to as the
{\em consumption rate} and {\em production rate} of the associated
input and output, respectively. If the source and sink actors of a
dataflow graph edge are SDF actors, then the edge is referred to as an
SDF edge, and if a dataflow graph consists of only SDF actors, and SDF
edges, the graph is referred to as an SDF graph.

For a dataflow graph edge $e$, $\ensuremath{\mathrm{src}(e)}$ and
$\ensuremath{\mathrm{snk}(e)}$, denote its source and sink actors, and
if $e$ is an SDF edge, then $\ensuremath{\mathrm{prd}(e)}$ denotes the
production rate of the output port of $\ensuremath{\mathrm{src}(e)}$
that is connected to $e$, and similarly,
$\ensuremath{\mathrm{cns}(e)}$ denotes the consumption rate of the
input port of $\ensuremath{\mathrm{snk}(e)}$ that is connected to $e$.
 
A {\em static schedule} for a dataflow graph $G$ is a sequence of
actors in $G$ that represents the order in which actors are fired
during an execution of $G$.

Usually, production and consumption information --- in particular, the
number of tokens produced and consumed (production/consumption {\em
  volume}) --- by individual firings is characterized in terms of
individual input and output ports so that each port of an actor can in
general have a different production or consumption volume
characterization. Such characterizations can involve constant values
as in SDF~\citep{lee1987x2} (as described above); periodic patterns of
constant values, as in cyclo-static dataflow
(CSDF)~\citep{bils1996x1}; or more complex forms that are
data-dependent (e.g., see~\citep{buck1993x1, bhat2000x4, murt2002x1,
  mcal2004x1, plis2008x1}). A meta-modeling technique called
parameterized dataflow (PDF) allows limited forms of dynamic
behavior~\citep{bhat2000x4} in terms of run-time changes to dataflow
graph parameters. The Boolean dataflow (BDF)~\citep{buck1993x1} and
core functional dataflow (CFDF)~\citep{plis2008x1} models are highly
expressive (Turing complete) dynamic dataflow models. We have
explained SDF, CSDF, and PDF models in greater detail later in this
section.

Apart from DIF, which we have mentioned earlier, there are various
existing design tools with their semantic foundations in dataflow
modeling, such as Ptolemy~\citep{pino1995x1},
LabVIEW~\citep{john1997x1}, StreamIt~\citep{thie2002x1},
CAL~\citep{eker2003x2}, PeaCE~\citep{kwon2004x1},
Compaan/Laura~\citep{stef2004x1}, and
SysteMoc~\citep{haub2007x1}. Dataflow-oriented DSP design tools
typically allow high-level application specification, software
simulation, and possibly synthesis for hardware or software
implementation~\citep{bhat2010x1}.

\subsubsection{Synchronous Dataflow}
\label{sec:sdf}

\begin{figure}[hb]
\centering
\noindent\includegraphics[width=6cm]{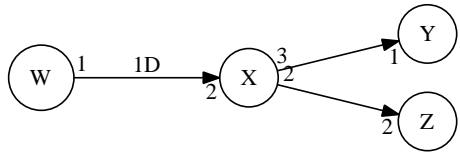}
\caption{An SDF graph.}
\label{fig:back-sdf}
\end{figure}

An SDF graph is characterized by its compile-time predictability
through the statically known consumption and production rates, as
defined above. Fig.~\ref{fig:back-sdf} shows a simple SDF graph having
actors \texttt{W}, \texttt{X}, \texttt{Y}, and \texttt{Z} (shown as
circles or vertices of the graph). Each edge (an arrow in the figure
connecting a pair of actors) is annotated with the number of tokens
produced on it by the source actor and that consumed from it by the
sink actor during every invocation of the source and sink actors,
respectively. For example, actor \texttt{X} can be fired when there
are at least two tokens on its input. Whenever actor \texttt{X} is
fired, it consumes two tokens from its input buffer, and produces
three tokens onto the output buffer connected to \texttt{Y} and two
tokens onto the output buffer connected to \texttt{Z}.

\subsubsection{Cyclo-static Dataflow}
\label{sec:csdf}

\begin{figure}[hb]
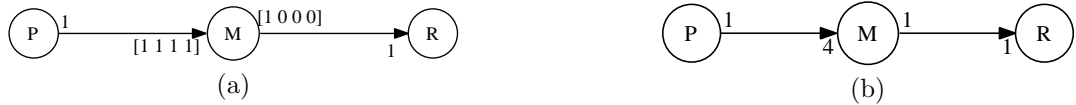

\begin{minipage}[hb]{0.48\linewidth}
\noindent\includegraphics[width=6cm]{fig1a.epsi}
\centering \centerline{(a)}
\end{minipage}
\hfill
\begin{minipage}[hb]{0.48\linewidth}
\noindent\includegraphics[width=5.5cm]{fig1b.epsi}
\centering \centerline{(b)}
\end{minipage}
\caption{An application graph with a simple decimator actor \texttt{M}
  using the (a) CSDF, and (b) SDF models. Actor \texttt{M} is a
  decimator with a {\em decimation factor} of $4$.}
\label{fig:back-csdf}
\end{figure}

Many signal processing applications involve behaviors in which
production and consumption rates may change during run-time. In some
cases, these changes may, however, be known at compile-time. For
example, consider the CSDF graph shown in Fig.~\ref{fig:back-csdf}(a),
which has a {\em decimator} actor \texttt{M} in it. This actor
consumes one token from its input on each invocation, but produces a
token onto its output only on every fourth invocation. This behavior
has been depicted using the varying production volumes denoted by $[1
  \, 0 \, 0 \, 0]$. The numbers of tokens produced by the decimator
\texttt{M} follow this cyclic pattern with a period of $4$.  This
sequence of varying production volumes, though not leading to constant
output rates like an SDF actor, is still completely deterministic and
known at the compile-time. This kind of dataflow behavior, where
actors exhibit token production and consumption volumes (in terms of
tokens per firing on specific actor ports) that are either constant or
expressible as cyclic sequences of constant volumes, is referred to as
CSDF. Thus, CSDF can be viewed as a generalization of SDF in which
token production and consumption volumes may be different across
different firings of an actor, but follow cyclic patterns that are
completely specified at the compile-time.

We refer readers to ~\citep{bils1996x1} for more details on the CSDF
model. As shown in Fig.~\ref{fig:back-csdf}(a) and
Fig.~\ref{fig:back-csdf}(b), it may be possible to transform a CSDF
actor into an SDF actor. In general, when feedback loops are present
in a dataflow graph, such a transformation may introduce deadlock, and
therefore should be attempted with caution.  Such a transformation,
when admissible (not leading to deadlock), generally has trade-offs in
terms of relevant metrics including latency, throughput, and code
size. More detailed comparisons between the SDF and CSDF models of
computation are presented in~\citep{park1995x2}
and~\citep{bhat2000x6}.

\subsubsection{Parameterized Dataflow}
\label{sec:pdf} 

Though CSDF provides enhanced expressive power compared to SDF, it is
still unable to specify patterns in token consumption and production
volumes that are not fully known at compile time. A meta-modeling
technique called PDF has been proposed to represent certain kinds of
dataflow application dynamics~\citep{bhat2000x4}. This model can be
used with any arbitrary dataflow graph format that has a well-defined
notion of a {\em schedule iteration}. For example, the PDF meta-model,
when combined with an underlying SDF model, results in the PSDF
(parameterized synchronous dataflow) model. A PSDF graph behaves like
an SDF graph during one schedule iteration, but can assume different
configurations across different schedule iterations.

The PDF meta-model supports semantic and syntactic
hierarchy. Syntactic hierarchy is used, as in other forms of dataflow,
to decompose complex designs in terms of smaller components. On the
other hand, semantic hierarchy in PDF is used to apply specific
features in the meta-model that are associated with dynamic parameter
reconfiguration.  A hierarchical actor that encapsulates such semantic
hierarchy in PDF is called a {\em PDF subsystem}. A PDF subsystem in
turn has three underlying graphs called the {\em init}, {\em subinit},
and {\em body} graphs, which interact with each other in structured
ways.  Intuitively, the init and subinit graphs can capture
data-dependent, dynamic behavior at certain points during the
execution of the graph and configure the body graph to adapt in useful
ways to such dynamics. Intuitively, the init graph is designed to
capture parameter configuration that is driven by higher, system-level
processing, while the subinit graph is designed to capture the
parameter changes occurring across different iterations of the
corresponding body graph. The init graph can be used to dynamically
configure parameters in the subinit graph, which, in general, executes
more frequently relative to the init graph.

\begin{figure}[hb]
\begin{minipage}[hb]{1.0\linewidth}
\centering
\noindent\includegraphics[width=10cm]{fig2a.epsi}
\centering \centerline{(a)}
\end{minipage}
\begin{minipage}[hb]{1.0\linewidth}
\centering
\noindent\includegraphics[width=10cm]{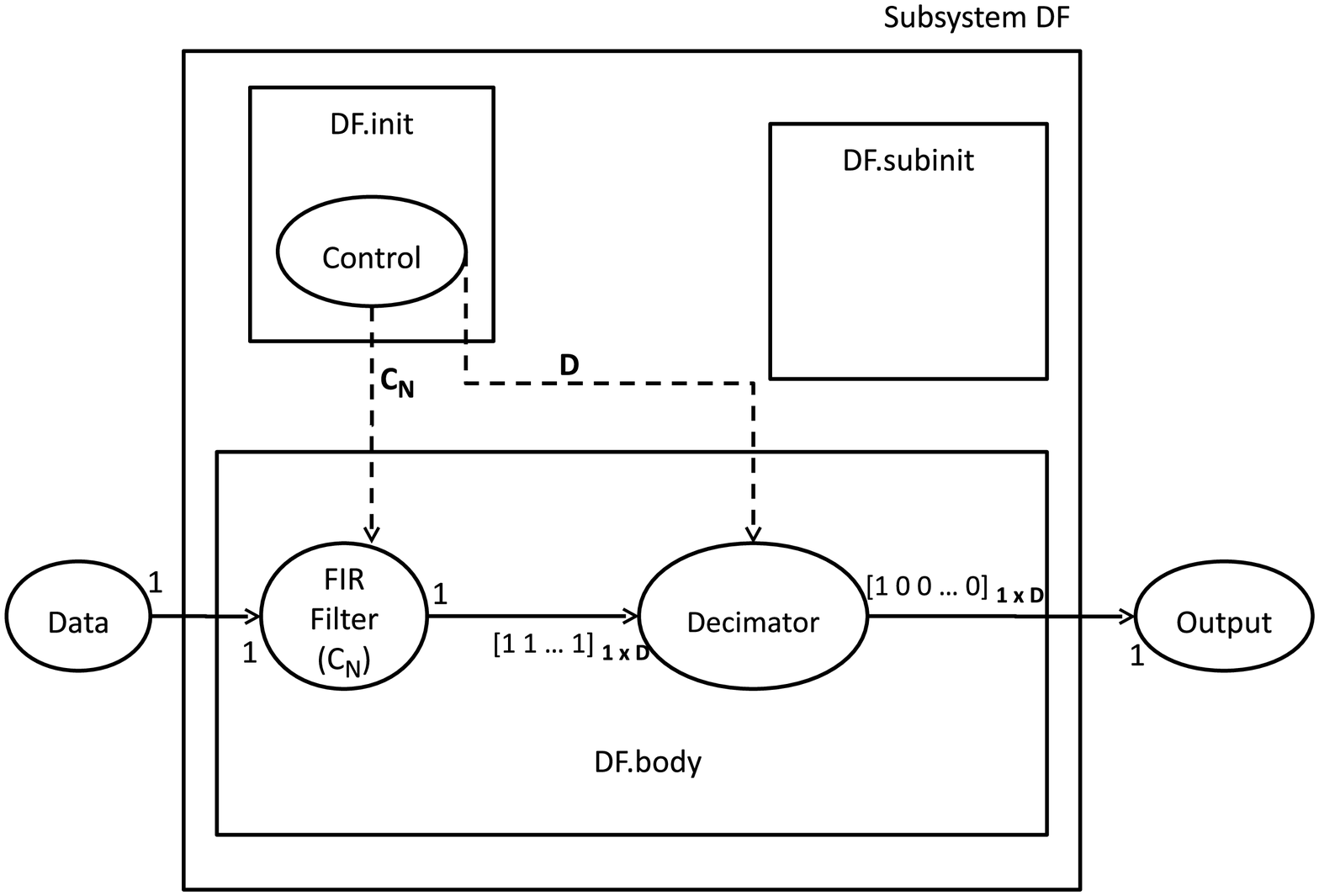}
\centering \centerline{(b)}
\end{minipage}
\caption{Modeling a parameterized decimation filter (\texttt{DF})
  application using PCSDF: (a) Application graph --- $C_N$ denotes a
  vector of FIR filter coefficients, and $D$ denotes a decimation
  factor, and (b) PCSDF representation.}
\label{fig:back-pcsdf}
\end{figure}

To further illustrate the PDF modeling technique, we consider the
application example shown in Fig.~\ref{fig:back-pcsdf}(a). This
example involves an FIR filter with filter taps or coefficients given
by $C_N = [c_0, c_1, \dots, c_{N-1}]$ followed by a decimator with a
tunable decimation factor of $D$. The values of $D$ and $C_N$ are set
either through a higher level system or user interface. We skip the
details of this mechanism for the sake of simplicity and
conciseness. Such behavior can be modeled using PDF with an underlying
CSDF model. Such a modeling approach is referred to as the {\em
  parameterized cyclo-static dataflow} ({\em PCSDF})
model~\citep{saha2006x2}. Fig.~\ref{fig:back-pcsdf}(b) shows one of
the possible PCSDF graphs corresponding to the application shown in
Fig.~\ref{fig:back-pcsdf}(a). The subsystem \texttt{DF} is a PCSDF
subsystem with its component graphs as shown in the figure. It can be
seen here that the \texttt{control} actor in the \texttt{DF.init}
graph of \texttt{DF subsystem} sets the required external and internal
parameters, $D$, and $C_N$, respectively. This actor models the
required parameter control through either a higher level system or
some form of user interface. In this particular case, the
\texttt{DF.subinit} graph is empty (in general, the init, subinit and
body graph do not all have to be used for a given subsystem).

The PCSDF model allows CSDF actors for which the cyclic patterns of
token production and consumption volumes can be parameterized in terms
of their {\em periods}, the actual numbers of tokens consumed or
produced in the cyclo-static sequences, or both. Intuitively, for a
given configuration of application parameters, a PCSDF graph behaves
as a CSDF graph. However, a PCSDF graph not only models all possible
parameter configurations in a given application but also describes how
they can be changed at run-time.

Such a model is of particular interest for modeling multirate DSP
systems that exhibit parameterizable sample rate conversions.  PCSDF
allows designers to systematically explore design spaces across
static, quasi-static, and dynamic implementation techniques. Here, by
{\em quasi-static} implementation techniques, we mean techniques where
relatively large portions of the associated software or hardware
structures are fixed at compile-time with minor adjustments allowed at
run-time (e.g., in response to changes in input data or operating
conditions). A variety of quasi-static dataflow techniques are
discussed, for example, in~\citep{bhat2010x1}.

\subsection{The Dataflow Interchange Format}
\label{sec:dif}
To describe dataflow applications for a wide range of DSP
applications, application developers can use {\em the DIF language},
which is a standard language founded in dataflow semantics and
tailored for DSP system design~\citep{hsu2005x2}. DIF provides an
integrated set of syntactic and semantic features that help promote
high-level modeling, analysis, and optimization of DSP applications
and their implementations without over-specification. From a dataflow
point of view, DIF is designed to describe mixed-grain graph
topologies and hierarchies as well as to specify dataflow-related and
actor-specific information. The dataflow semantic specification is
based on dataflow modeling theory and independent of any design tool.

\begin{figure}[hb]
\centering
\noindent\includegraphics[scale=0.8]{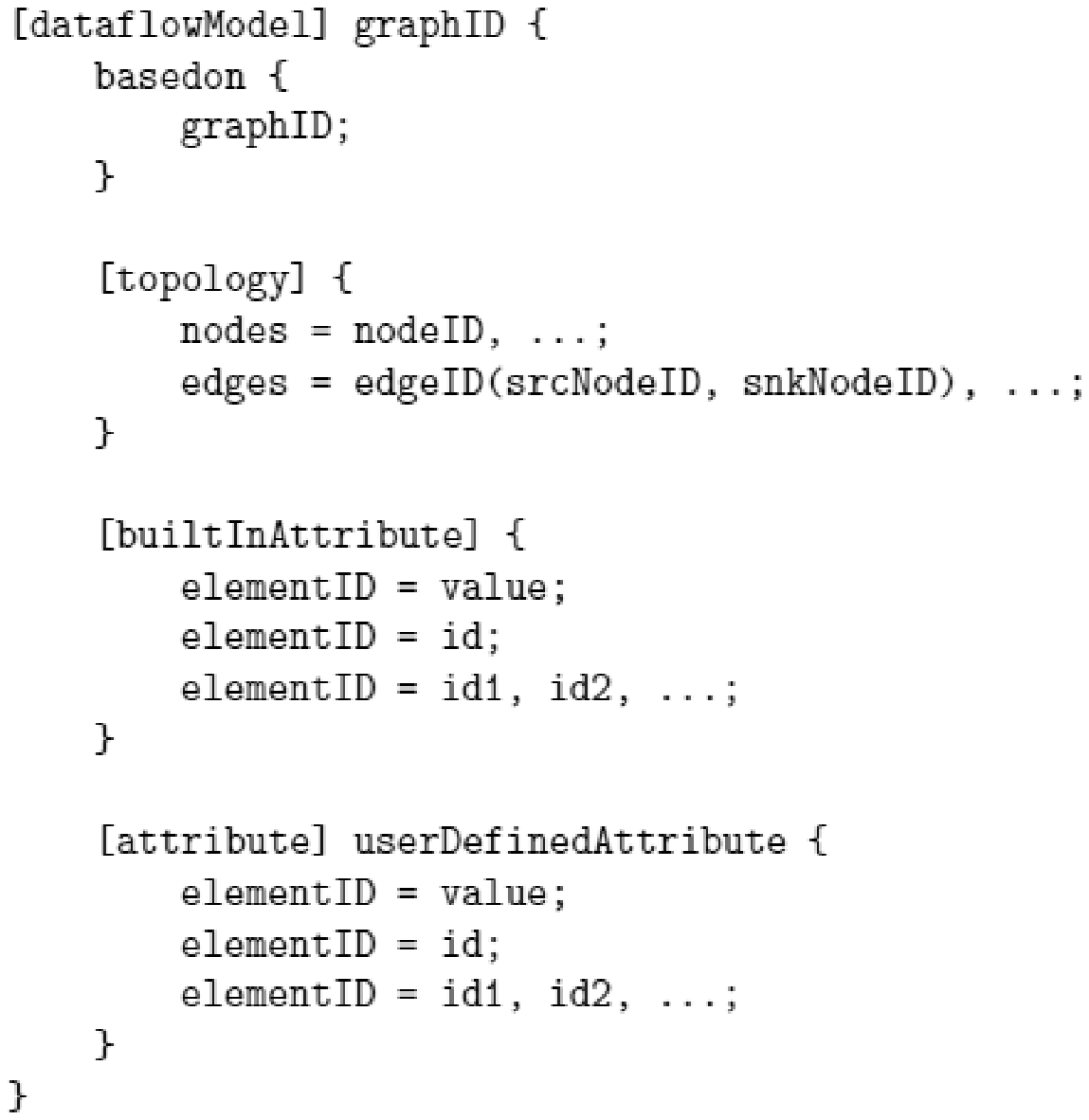}
\caption{The DIF language.}
\label{fig:dif}
\end{figure}

Fig.~\ref{fig:dif} illustrates some of the available constructs in the
DIF language along with the syntax used for application
specification. More details on the DIF language can be found
in~\citep{hsu2007x4}. The \texttt{topology} block of the specification
specifies the graph topology, which includes all of the \texttt{nodes}
and \texttt{edges} in the graph. DIF supports {\em built-in
  attributes} such as \texttt{interface}, \texttt{refinement},
\texttt{parameter}, and \texttt{actor}, which identify specifications
related to graph interfaces, hierarchical subsystems, dataflow
parameters, and actor configurations, respectively. DIF also allows
{\em user-defined attributes}, which have a similar syntax as built-in
attributes except that they need to be declared with the
\texttt{attribute} keyword.

The DIF language has been recently augmented with constructs for
supporting {\em topological patterns}~\citep{sane2010x3}. Topological
patterns allow concise specification of functional structures at the
dataflow graph (inter-actor) level. They can effectively represent
many of the flowgraph substructures that are pervasive in the DSP
application domain (e.g. {\tt chain}, {\tt ring}, {\tt butterfly},
etc.) to generate compact, scalable application representations. We
direct readers to~\citep{sane2010x3, sane2011x1} for more information
on the concept of topological patterns and how the DIF supports it.

\begin{figure}[hb]
\begin{minipage}[hb]{1.0\linewidth}
\centering
\noindent\includegraphics[scale=0.5]{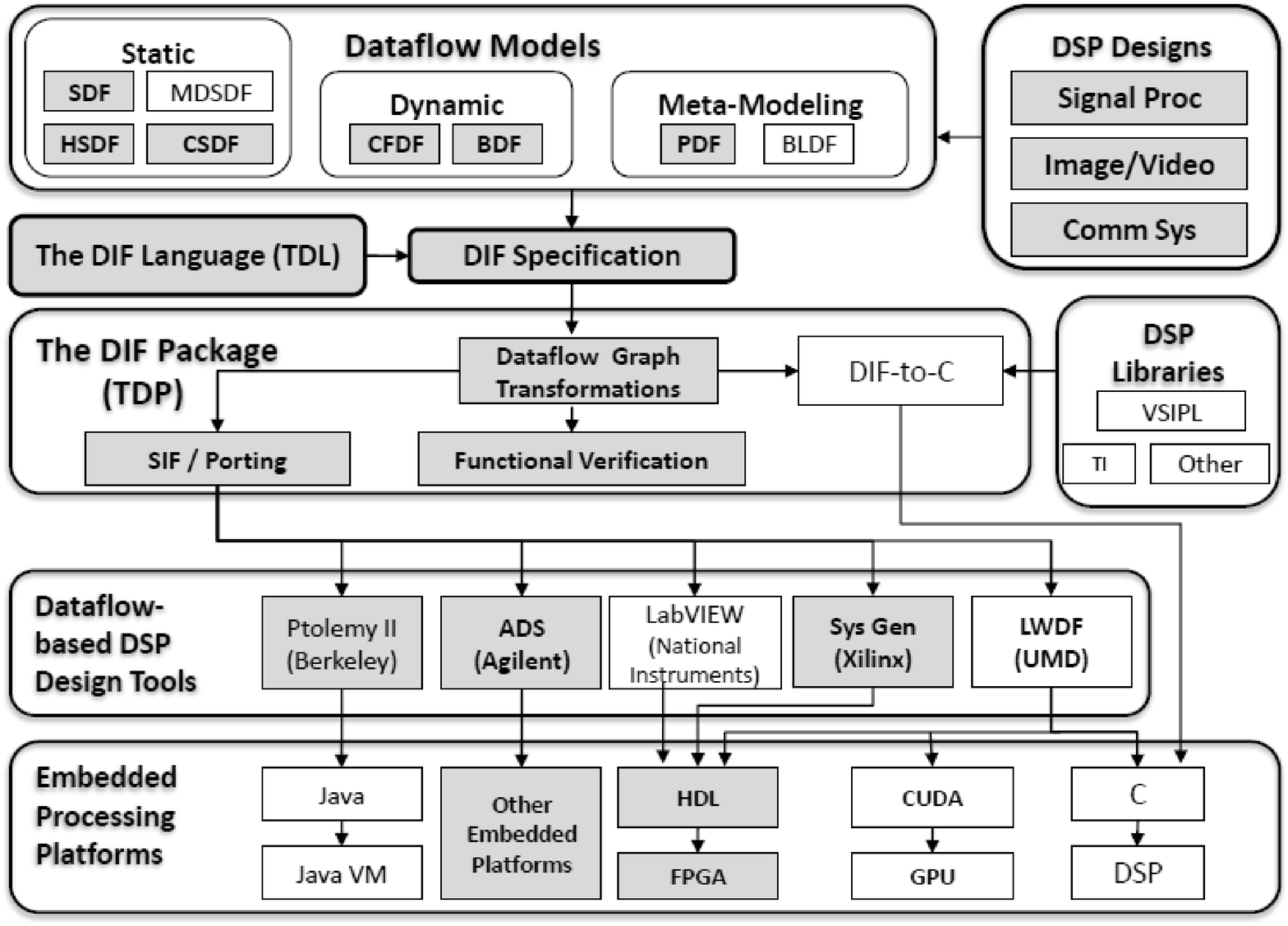}
\end{minipage}
\caption{The DIF Package.}
\label{fig:tdp}
\end{figure}

To facilitate use of the DIF language, {\em the DIF package} ({\em
  TDP}) has been built (see Fig.~\ref{fig:tdp}). Along with the
ability to transform DIF descriptions into manipulable internal
representations, TDP contains graph utilities, optimization engines,
verification techniques, a comprehensive functional simulation
framework, and a software synthesis framework for generating C
code~\citep{hsu2005x2, plis2008x1}. These facilities make TDP an
effective environment for modeling dataflow applications, providing
interoperability with other design environments, and developing and
experimenting with new tools and dataflow techniques. Beyond these
features, DIF is also suitable as a design environment for
implementing dataflow-based application representations.  Describing
an application graph is done by listing nodes (actors) and edges, and
then annotating dataflow specific information as well as other
(non-dataflow) kinds of relevant information associated with actors,
edges, and design subsystems.

The framework in DIF for simulation and functional verification of
applications, which is based on CFDF semantics, allows application
specifications in DIF to be used as executable references for rapid
system prototyping and developing further platform-specific
implementations. CFDF, which supports dynamic dataflow behaviors,
allows flexible and efficient prototyping of dataflow-based
application representations, and permits natural description of both
dynamic and static dataflow actors. More information on CFDF semantics
can be found in~\citep{plis2008x1}.

\subsection{Related Work}
\label{sec:work}
There exist high-end reusable, modular, scalable, and reconfigurable
FPGA platforms such as the {\em Berkeley Emulation Engine 2} ({\em
  BEE2})~\citep{chan2005x1}, IBOB~\citep{pars2006x1}, and
UniBoard~\citep{szom2011x1}, which have been introduced specifically
for DSP systems. These have been widely used for radio astronomy
applications. The BEE2 uses SDF as a unified computation model for
both the microprocessor and the reconfigurable fabric. It uses a
high-level block diagram design environment based on The Mathworks'
Simulink and the Xilinx System Generator (XSG). This design
environment, however, does not expose the underlying dataflow
model. In particular, the designer has little or no scope to make use
of the underlying dataflow model for experimentation (as mentioned
earlier in Section~\ref{sec:intro}). Also, the SDF model used for
programming the BEE2 is a static dataflow model in that all the
dataflow information is available at compile-time (i.e., before
executing or running the application). Though this feature provides
maximal compile-time predictability, it has limited expressive
power. It does not allow for data-dependent, dynamic behavior, which
is exhibited by many modern DSP applications, such as the TDD
application introduced in Section~\ref{sec:tdd-app}
(see~\citep{bhat2010x1} for more examples of such applications). Other
forms of dataflow models that can capture more application dynamics
with acceptable levels of compile-time predictability may better
exploit the features offered by platforms such as the BEE2. We should,
however, mention that the CASPER DSP library offers a software
register block that can provide limited parameterization in the
design. We have used this block extensively in our TDD design.

There are some other FPGA design solutions and tool flows available
(e.g., those from Nallatech~\citep{nallurl}, and
Lyrtech~\citep{lyrturl}). These, however, are commercial tools and do
not provide open-source DSP software libraries like the CASPER. Also,
CASPER tools support most of the Xilinx FPGA devices unlike these
other commercial tools.

Model based approaches for designing large scale signal processing
systems with a focus on radio telescopes have been previously studied
(e.g., see~\citep{alli2004x1, lema2006x1, lema2008x1}). Several
frameworks have been proposed for model based, high-level abstractions
of architectures along with performance/cost estimation methods to
guide the designer throughout the development cycle (see
~\citep{alli2004x1}). However, the focus of these approaches has been
on architecture exploration. There have also been attempts to derive
implementation-level specifications starting from system-level
specifications by segregating signal processing and control flow
(see~\citep{lee2011x1} for more information on control flow) into an
application specification and architecture specification, respectively
(see ~\citep{lema2006x1, lema2008x1}). However, the choice of models
of computation has been made primarily from control flow
considerations rather than dataflow considerations. These approaches,
though relevant, do not specifically address the issue of high-level
application specification for platform-independent prototyping and use
of models of computation for abstraction of heterogeneous or hybrid
dataflow behaviors. This issue is critical to efficient prototyping of
high performance signal processing applications, which are typically
dataflow dominated, and include increasing levels of dynamic dataflow
behavior (e.g., see~\citep{bhat2010x1}).

We address this issue using the CFDF model with underlying PSDF or
PCSDF behavior and using it for system prototyping. We then show how
platform-independent specifications based on this modeling technique
can be used to efficiently develop platform-specific implementations.

\section{Dataflow-based Design and Implementation of a TDD}
\label{sec:design-flow}
\begin{figure}[hb]
\centering
\noindent\includegraphics[scale=0.5]{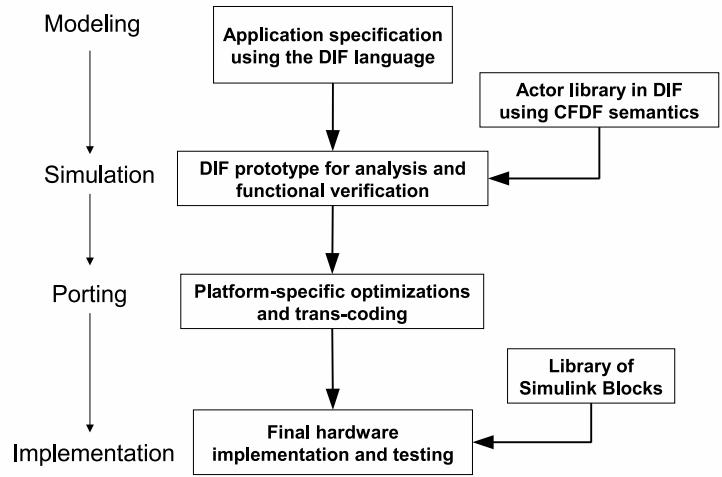}
\caption{Dataflow-based approach for design and implementation of a TDD.}
\label{fig:design-flow}
\end{figure}

We propose an approach for design and implementation of a TDD based on
the dataflow formalisms discussed in Section~\ref{sec:df} along with
relevant capabilities of the DIF tool described in
Section~\ref{sec:dif}. Fig.~\ref{fig:design-flow} gives an overview of
our dataflow based approach, which we now describe.

\subsection {Modeling and Prototyping using DIF}
\label{sec:prototype}

We start with an application specification that describes the DSP
algorithm under consideration (in this case, the TDD) along with
proper input and output interfaces. The application is specified using
the DIF language. This DIF specification consists of topological
information about the dataflow graph --- interconnections between the
actors along with input and output interfaces. The DIF specification
is a platform-independent, high-level application specification. The
specification can be used, for example, to simulate the application,
given the library of actors from which the specification is
constructed.

Depending upon the application under consideration, the designer can
select among a variety of dataflow models of computation in DIF to
effectively capture relevant aspects of the application dynamics. It
should be noted that the designer does not always need to specify the
model in advance. The CFDF model can be used to describe individual
modules (actors) in the application, and the DIF package can analyze
the CFDF representation (CFDF modes, to be specific) of the actors, as
specified by the designer through the actor code, and annotate the
actors with additional dataflow information using various techniques
for identifying specialized forms of dataflow behavior (e.g.,
see~\citep{plis2010x3}).  This step requires the functionality of
individual actors to be specified in CFDF semantics. The designer can
use the existing blocks from the Java actor library in DIF or develop
his or her own library of CFDF actors.

In terms of tunability, the key components of the TDD as seen from
Fig.~\ref{fig:tdd} are the tunable FIR filter, and decimation filter
blocks. The tunable decimation filter (TDF) block is of particular
interest, considering that it is the only multirate block in the
system. Its behavior resembles that of the one described in
Section~\ref{sec:pdf}. In view of this, we have identified PSDF and
PCSDF as candidate dataflow models for efficient implementation of the
targeted TDD system. For this system, we have to take into account the
multiple inputs and outputs to actors, as mentioned in
Section~\ref{sec:tdd-app}.

\begin{figure}[hb]
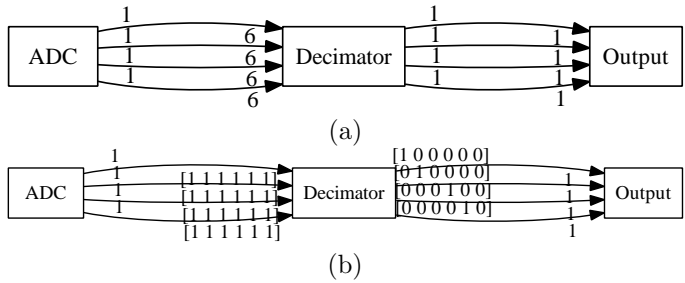

\begin{minipage}[hb]{1.0\linewidth}
\noindent\includegraphics[width=9cm]{fig10a.epsi}
\centering \centerline{(a)}
\end{minipage}
\begin{minipage}[hb]{1.0\linewidth}
\noindent\includegraphics[width=9cm]{fig10b.epsi}
\centering \centerline{(b)}
\end{minipage}
\caption{Dataflow behavior of a \texttt{Decimator} actor with $4$ inputs and
  outputs for a {\em decimation factor} of $6$ using (a) SDF, and (b)
  CSDF models.}
\label{fig:tdd-df}
\end{figure}

To illustrate details of the dataflow behavior of a decimator actor
based on such specifications, we have shown one such decimator actor
with $4$ inputs and outputs, and having a decimation factor of $6$ in
Fig.~\ref{fig:tdd-df}(a) and Fig.~\ref{fig:tdd-df}(b). The decimator
simultaneously receives $4$ consecutive samples from its $4$
inputs. It outputs every sixth input sample starting with the first
input sample. Each of these output samples appears on a successive
output of the decimator.

For the sake of simplicity and clarity, we have excluded the other
single rate blocks from the application graphs in these figures. In
our implementation, we extend this behavior for an actor with $8$
inputs and outputs. We have created a DIF prototype using PSDF and
PCSDF as underlying models for equivalent CFDF representation of actor
blocks. We have also developed a Java library of actors in DIF
adhering to CFDF semantics for all of the blocks.

We then used DIF for software prototyping, analysis, and functional
simulation.  The DIF package uses the DIF specification to generate an
intermediate graph representation, which can then be used as an input
for further graph transformations including a {\em scheduling}
transformation, which determines the schedule for an application.
Here, by a {\em schedule}, we mean the assignment of actors to
processing resources, and the execution ordering of actors that share
the same resource. The functional simulation capabilities provided in
DIF can be used to analyze and estimate buffer requirements in terms
of the numbers of tokens accumulated on the buffers that correspond to
dataflow graph edges. This provides an estimate of total memory
requirements as well as specifications for individual buffers when
porting the application to the targeted implementation platform.

\begin{figure}[hb]
\centering
\noindent\includegraphics[width=9cm]{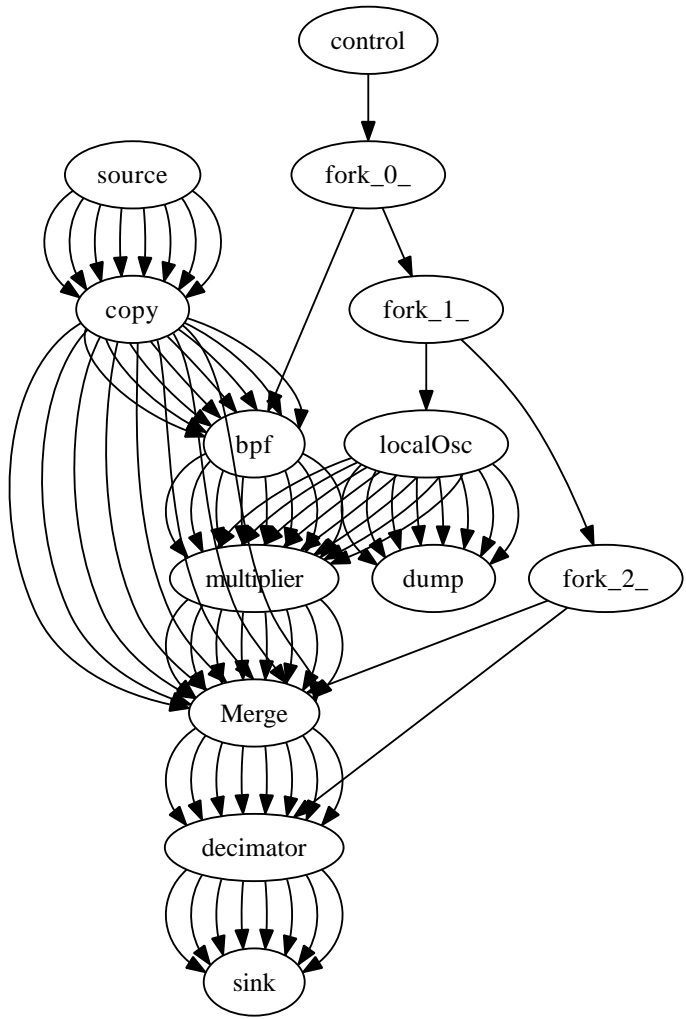}
\caption{TDD application graph generated using DIF.}
\label{fig:tdd-dif-graph}
\end{figure}

\begin{figure}[hb]
\centering
\noindent\includegraphics[scale=0.7]{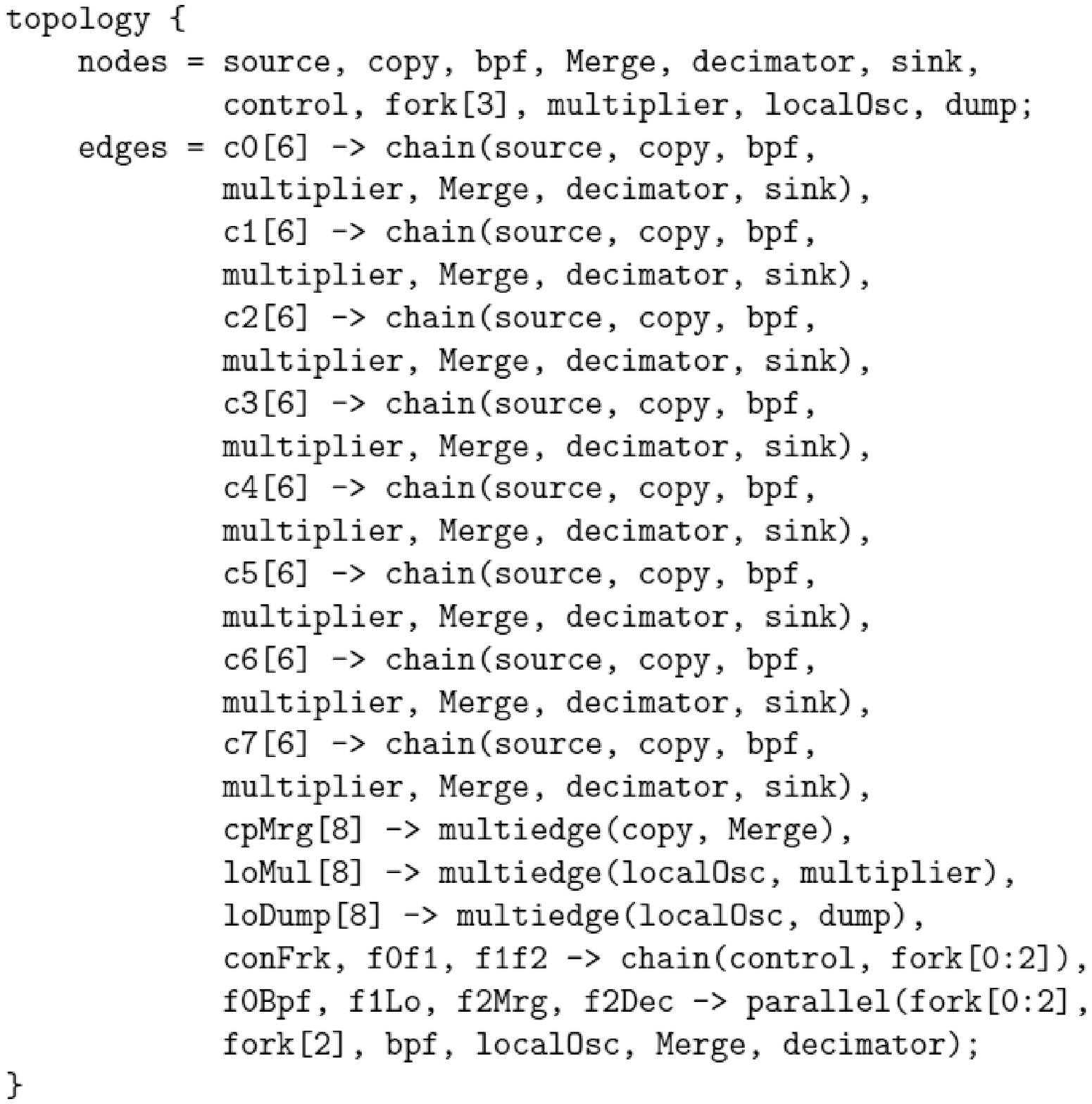}
\caption{Partial DIF specification --- {\tt topology} block --- 
for the TDD application graph using topological patterns.}
\label{fig:tdd-spec-2}
\end{figure}

Fig.~\ref{fig:tdd-dif-graph} shows the TDD application graph generated
using DIF. This is based on the TDD block diagram shown in
Fig.~\ref{fig:tdd} with addition of some actors that handle parameter
configuration for the actors. We discard one of the two sets of
outputs (more specifically, {\em sine} output) of
the \texttt{localOsc} actor as we have employed a real mixer in our
design. The complexity of the graph, which is increased due to
multiple parallel edges between two actors, can easily be captured
through a DIF specification that makes use of topological patterns. We
have shown one of the possible specifications of the graph topology in
DIF using topological patterns in Fig.~\ref{fig:tdd-spec-2}.

\begin{figure}[hb]
\begin{minipage}[t]{1.0\linewidth}
\centering 
\noindent\includegraphics[scale=0.45]{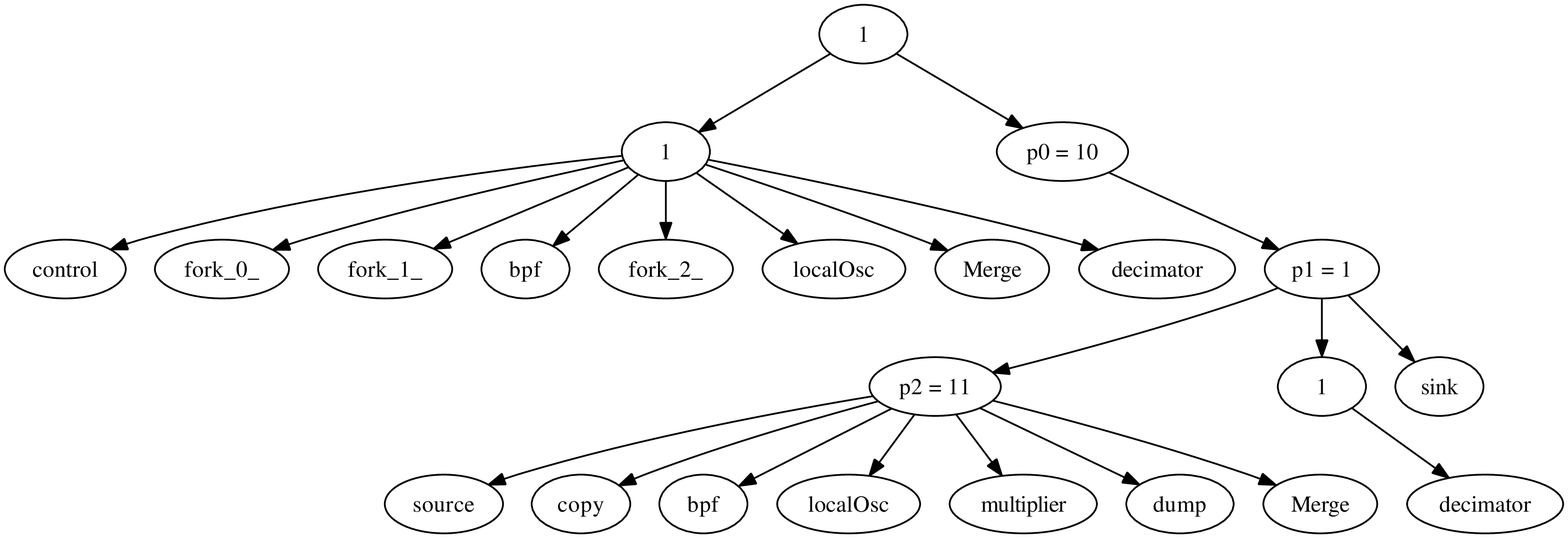}
\centering \centerline{(a)}
\end{minipage}
\medskip
\begin{minipage}[t]{1.0\linewidth}
\centering
\noindent\includegraphics[scale=0.45]{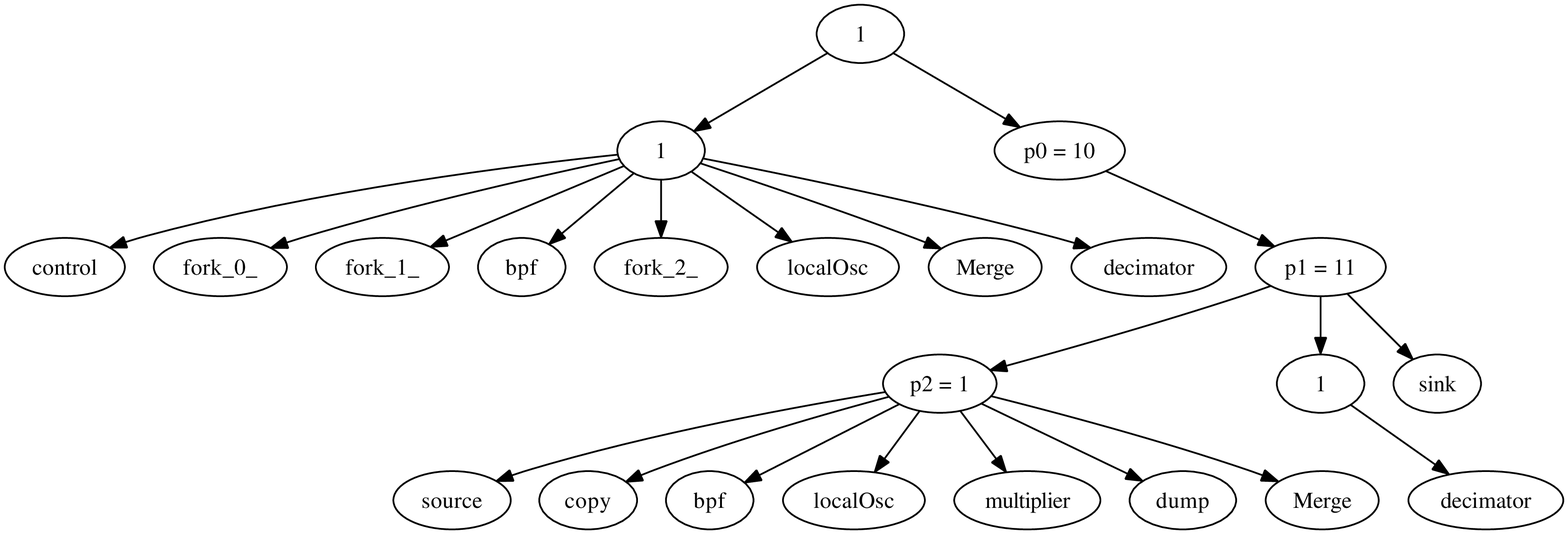}
\centering \centerline{(b)}
\end{minipage}
\caption{PLSs for the TDD application configured for a decimation factor 
of $11$, and \texttt{decimator} actor employing the (a) PSDF and (b)
PCSDF models of computation.}
\label{fig:tdd-pls}
\end{figure}

For our design, we have used parameterized looped schedules
(PLSs)~\citep{ko2007x1} for PSDF and PCSDF models to determine the
total buffer requirements. Using the TDD specification, we construct
PLSs for the TDD application. Fig.~\ref{fig:tdd-pls}(a) shows a PLS
for a TDD application, where the \texttt{decimator} actor has the
underlying SDF model, while Fig.~\ref{fig:tdd-pls}(b) shows one in
which the \texttt{decimator} actor employs the CSDF model. We have
used the {\em generalized schedule tree} ({\em GST}) representation
for the PLSs~\citep{ko2007x1}. An internal node of a GST denotes a
loop count, while a leaf node represents an actor. The execution of a
schedule involves traversing the GST in a depth-first manner, and
during this traversal, the sub-schedule rooted at any internal node is
executed as many times as specified by the loop count of that node. As
annotated in these GSTs, loop counts \texttt{p0}, \texttt{p1},
and \texttt{p2} are parameterizable. The loop count \texttt{p0} is set
to a user-specified number of iterations, while the loop
counts \texttt{p1} and \texttt{p2} are tuned based upon the decimation
factor as well as the underlying dataflow model for the
\texttt{decimator}. Fig.~\ref{fig:tdd-pls}(a) and (b), in particular, show
values of the parameterizable loop counts set for a \texttt{decimator}
with a decimation factor of $11$. This PLS can be viewed as providing
CFDF-based execution for the given PDF-based actor specification
model.

Table~\ref{tbl:buffer} shows the total buffer requirements using PLSs
shown in Fig.~\ref{fig:tdd-pls}(a) and (b) for various configurations
of decimation factors. Note that for a given configuration (setting of
graph parameters), a PSDF or PCSDF graph behaves like an SDF or CSDF
graph, respectively. It can be seen that for the SDF model, the total
buffer requirements vary with the decimation factor, and this is due
to input buffers to the TDD block that need to accumulate varying
numbers of tokens. Thus, employing the PSDF model will require tuning
buffer sizes for different decimation factors if one wants to provide
for optimized buffer sizes in terms of graph parameters.

We have used the CASPER tool flow for developing our platform-specific
implementation as explained later in Section~\ref{sec:casper}.  This
implementation is targeted to an FPGA. Our objective here is to
support tuning the decimation factor without regenerating hardware
code. A dataflow buffer can be implemented using a FIFO or dual-port
random access memory (RAM) block in the targeted FPGA device. The size
of the available FIFO block can be set to $2^n$, where $n \geq
1$. This gives limited control over setting the FIFO size, and may
increase the resource utilization. At the same time, tuning the sizes
of FIFO or dual-port RAM blocks is not possible during run-time. It is
in general possible to set the size of a FIFO or dual-port RAM block
to a maximum required value, and access only a part of it using a
tunable address counter during run-time. This, however, again may lead
to unnecessary increased resource utilization. The ADC output is of a
streaming nature (data is produced or consumed at every clock cycle
without any synchronization signal), as is the DSP subsystem
downstream of the TDD.

In order to achieve the throughput constraint imposed by the maximum
data rate of the ADC output stream, SDF buffers need to be pipelined,
which is not efficient using RAM blocks. Thus, we use the CSDF model,
which does not require tuning of dataflow buffer sizes to achieve the
maximum throughput constraint, as observed from our DIF-based
prototype.  The TDD generates a synchronization or enable signal
indicating a valid output data. This can be used as a clock to drive
the downstream DSP system.

We use our DIF prototype as a reference while integrating the design
with the current CASPER tool flow for the target implementation on the
IBOB. Section~\ref{sec:casper} further elaborates on this approach
along with implementation results.

\subsection{Integration with the CASPER Tool Flow}
\label{sec:casper}

The CASPER tool flow is based on the BEE\_XPS tool
flow~\citep{pars2006x1}. This tool flow requires that an application
be specified as a Simulink model using XSG~\citep{pars2006x1}. Since
there is no automated tool for transforming a DIF representation into
an equivalent Simulink model, porting the DIF specification to
Simulink/XSG requires manual transcoding of the DIF
specification. This also requires implementing parameterizable actor
blocks that are currently not available in the XSG, CASPER, or
BEE\_XPS libraries.

Each actor gets transformed into an equivalent functional XSG
block. For each of the Simulink actor blocks, we provide a
pre-synthesis parameterization that allows changing block parameters
before hardware synthesis (see~\citep{pars2007x1} for more details on
Simulink scripting). In order to implement our objective of tunability
--- post-synthesis parameterization --- we use the {\em software
register} mechanism in the BEE\_XPS library to specify parameters that
change during run-time (that is, after hardware code is generated, and
depending upon user requirements.)

Software registers can be accessed and set during run-time from the
TinyShell interface available for IBOB. This allows tuning TDD
parameters without re-synthesizing the hardware each time the
parameters change from the previous setting. Each block has an enable
input signal. Through systematic transformations, an application graph
in DIF can be converted into an equivalent Simulink/XSG model. We have
developed an interface software package using C programs, and Bash and
Python scripts to compute software register values for the required
TDD configuration, and set these values on the IBOB over a telnet
connection, which is used for remote access to the hardware platform
at NRAO.

On the targeted FPGA device, we have employed the NCO using dual-port
RAM blocks that are loaded with pre-computed sinusoidal signal values
of the required precision. Each of these dual-port RAM blocks is used
to simultaneously read sine and cosine values from both of its ports.
The oscillator frequency is set using a software register, and depends
upon the desired output signal band.

In our current implementation, the TDF block (see Fig.~\ref{fig:tdd})
can have up to $16$ filter taps. We have also implemented a tunable
FIR filter block, which does not decimate, shown in
Fig.~\ref{fig:tdd}. This block can have up to $8$ taps in our
implementation. These, again, are set using software
registers. Fig.~\ref{fig:schem}(b) shows the schematic of a TDF. As
shown in this figure, we have employed two filter banks (16-tap units)
inside our design of a TDF block that operate in tandem to allow
maximum throughput (that is, the maximum data rate of the ADC output
stream). Hence, our TDF block has $32$ multiplication operations. As
mentioned earlier, our TDF design employs a polyphase implementation
as described in~\citep{vaid1990x1}. The software computes the sequence
in which the input signals should be routed to an appropriate filter
tap for a given decimation factor. This information is then fed to the
signal routing scheme using software registers.

Table~\ref{tbl:tdd} shows results for the TDD implementation on the
IBOB using the Xilinx EDK 7.1.2. We have used this hardware platform
and tool for all of the experiments reported in the remainder of the
paper. Design $1$ shows some of the device utilization parameters for
a TDD that supports only baseband modes. This design does not include
the tunable FIR filter, NCO, and mixer blocks shown in
Fig.~\ref{fig:tdd}. Design $2$ is based on the block diagram of a TDD
shown in Fig.~\ref{fig:tdd}.  As evaluation metrics for hardware cost,
we have used the utilization of FPGA slices, $4$-input look-up tables
(LUTs), and block RAM units, and the number of embedded
multipliers. Note that neither of these two designs use any of the
available embedded multipliers for multiplication. Designs $3$ and $4$
are modified versions of designs $1$ and $2$, respectively, in that
they employ embedded $18 \times 18$ multipliers. It can be seen that
using embedded multipliers does not provide significant improvements
in hardware cost. We observe that use of embedded multipliers, in
fact, needs to be accompanied by addition of extra latency in the
design to achieve timing closure. We have been able to achieve maximum
throughput using an implementation based on the PCSDF model.

\subsection{Platform-specific Analysis using DIF}
\label{sec:tool-development}

It is common to go back and forth between a high-level prototype and a
corresponding platform-specific implementation while designing an
embedded DSP system. Such alternation in design phases is common, for
example, when one is developing a platform-specific library or tool
flow. In support of such a design methodology, it is desirable for a
high-level design tool to support platform-specific analysis. This can
be achieved by annotating the high-level application specification
with platform-specific implementation parameters, which are derived
through device data sheets, experimentation or some combination of
both.

DIF supports specifying user-defined actor parameters. We use this
feature in DIF to annotate actors with two relevant implementation
parameters --- the latency constraint, and number of embedded
multipliers. This allows estimating results based on the DIF prototype
itself instead of determining them from the constructed design, which
is generally time consuming. We have verified the accuracy of metrics
estimated by our DIF model compared with actual hardware synthesis
results that are shown in Table~\ref{tbl:tdd}.

Developers of tool flows and DSP libraries can profile their library
blocks to determine a wide variety of platform-specific implementation
parameters. DIF can use such information to estimate implementation
parameters at a high-level of abstraction, and earlier in the design
cycle to help efficiently prune segments of the design space. Support
for estimation of various platform-specific resources for different
platforms is beyond the scope of this paper. It is, however, an
important direction toward developing alternative model based design
flows and open access tool flows for astronomical DSP solutions.

\subsection{Exploring Implementation Trade-offs between TDD and FDD Designs}
\label{sec:trade-off}
One of the motivations for the work presented in this paper has been
to develop library blocks needed for a TDD using Xilinx LogicCore and
CASPER library blocks. The current CASPER DSP library provides a
decimator (see Fig.~\ref{fig:schem}(a)) that supports decimation
factors that are powers of $2$. The decimation factor as well as the
filter coefficients of the FIR filter are not tunable after the
hardware code is generated. Our design provides flexibility with not
only the decimation factor but also the filter coefficients through
the use of software registers, as explained earlier. The FDD designs,
though not tunable, have lower hardware cost in terms of device
utilization. Table~\ref{tbl:trade-off} provides a summary of some of
the hardware utilization parameters for the FDD designs. These designs
have also been implemented on a CASPER IBOB. The decimation factor of
$10$ has been achieved by first interpolating the input by a factor of
$80$, and then decimating it by a factor of $8$. Comparison between
the results in this table and those in Table~\ref{tbl:tdd} clearly
highlights the trade-off between design flexibility and hardware
cost. Using the model-based approach presented in this paper, the
designer can effectively explore this trade-off based on the given
design requirements.

\subsection{TDD and FDD for Multistage Downconversion}
\label{sec:cascading}
\begin{figure}[hb]
\centering
\noindent\includegraphics[scale=0.6]{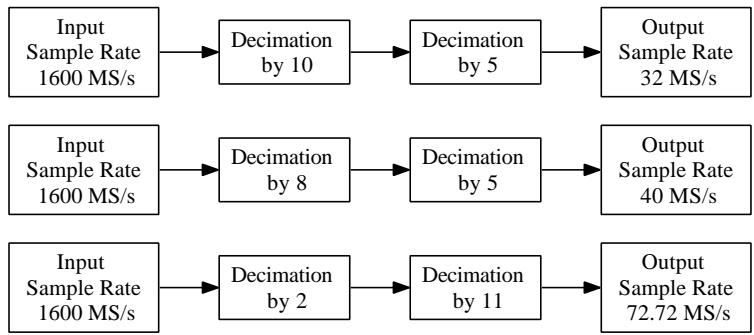}
\caption{Two-stage digital downconversion.}
\label{fig:cascading}
\end{figure}

Though our TDD design supports limited decimation factors (integer
factors between $5$ and $12$), its usage is not limited to these
factors. It can be readily scaled and applied to achieve other
decimation factors by cascading multiple TDF
blocks. Fig.~\ref{fig:cascading} shows some of the possible
input/output sampling rate relations that can be achieved by such use
of cascaded TDF blocks. Design $1$ in Table~\ref{tbl:cascading}
employs cascaded TDF blocks, while design $2$ in
Table~\ref{tbl:cascading} employs cascaded fixed-configuration
decimation filter (FDF) blocks. Both of these designs have been
developed to demonstrate multistage downconversion for a baseband
signal and hence, do not employ mixers. It is possible to extend these
designs to include a mixer to allow all possible narrow band outputs
and not just the baseband output. For all of the designs in this table
that use one or more TDF blocks, the TDF block employs dedicated
embedded multipliers.

In this light, we further explore the trade-off between the low
hardware cost of FDD designs and flexibility offered by TDD designs by
examining a design consisting of an FDF block followed by a TDF block
(designs $3$ and $4$ in Table~\ref{tbl:cascading}). These designs
provide limited tunable decimation factors compared to design $1$, but
also have lower hardware cost in terms of device utilization.

\section{Summary and Conclusions}
\label{sec:conclusion}
We have proposed a dataflow-based approach for prototyping radio
astronomy DSP systems. We have used a dataflow-based high-level
application model that provides a platform-independent specification,
and assistance in functional verification and important resource
estimation tasks. This can prove effective in reducing the development
cycle and faster deployment of DSP systems across various target
platforms. We have employed this approach to methodically develop a
TDD based DSP backend design. Our TDD implementation is targeted to
the CASPER FPGA board, called IBOB, and supports tuning narrow band
modes without the need for regenerating hardware code. We have also
explored the trade-off between the low hardware cost for FDD designs
and the flexibility offered by TDD designs. This trade-off has also
been highlighted in the context of designs employing a two-stage
downconversion scheme. A designer can explore this design space to
best meet the application requirements. Expanding on our work to
integrate TDDs with ongoing development of spectrometer designs at the
NRAO on the latest CASPER hardware is a natural extension of the work
presented in this paper.

There is a growing interest in the radio astronomy community to have
open-access and portable astronomical signal processing
solutions. Currently, this is constrained by proprietary commercial
tools targeted for specific platforms. We have also relied on these
tools, mainly for hardware synthesis and code generation, in our
work. In this context, it is of interest to have high-level
application description languages with semantic foundations in models
of computation, and the corresponding design tools for efficient
specification, simulation, functional verification, and
synthesis. Developing model based, platform-specific libraries, and
devising techniques for automatic code generation from high-level
representations, such as those in DIF, specifically for the radio
astronomy domain is an important direction for future research.


%
%
%
%
%
%
%

\begin{acknowledgments}
This research was sponsored in part by the National Radio Astronomy
Observatory, Austrian Marshall Plan Foundation, and National Science
Foundation (grant AGS-0959761 to New Jersey Institute of
Technology). We acknowledge with thanks the contributions of Shilpa
Bollineni, Srikanth Bussa, Randy McCullough, Scott Ransom, and Jason
Ray of the National Radio Astronomy Observatory. The National Radio
Astronomy Observatory is a facility of the National Science Foundation
operated under cooperative agreement by Associated Universities, Inc.
\end{acknowledgments}

%
%
%
%
%
%
%
%
%

\clearpage

\bibliographystyle{agu08} 
\bibliography{refs}









%

%
%

\end{article}




%
%
%
%
%
%
%


\clearpage


\clearpage

\begin{table}
\caption{Total buffer requirements from a DIF prototype for different
  decimation factors using parameterized looped schedules.}
\begin{tabular}{l c c c c c c c c c c}
\hline
Decimation Factor & & 5 & 6 & 7 & 8 & 9 & 10 & 11 & 12 \\ \hline
Total buffer requirements & SDF & 132 & 140 & 148 & 156 & 164 & 172 & 180 & 188 \\ 
(Number of tokens) & CSDF & 100 & 100 & 100 & 100 & 100 & 100 & 100 & 100 \\
\hline
\end{tabular}
\label{tbl:buffer}
\end{table}

\begin{table}
\caption{Implementation summary for TDD designs. In all the designs
  below, the input bandwidth is $800$~MHz, and decimation factor, $D$,
  is tunable such that $5 \leq D \leq 12$.}
\begin{tabular}{l c c c c}
\hline
Parameter  & Design 1 & Design 2 & Design 3 & Design 4 \\ \hline
Mixer & No & Yes & No & Yes  \\
Latency (ns) & 65 & 150 & 85 & 190 \\
FPGA slices (Out of 23616) & 12234 (52\%) & 13315 (56\%) & 12322 (52\%) & 14232 (60\%) \\
4 input LUTs (Out of 47232) & 14139 (29\%) & 16123 (34\%) & 12123 (25\%) & 15035 (31\%)  \\
Block RAMs (Out of 232) & 41 (17\%) & 48 (20\%) & 41 (17\%) & 48 (20\%) \\
$18 \times 18$ Multipliers (Out of 232) & --- & --- & 32 (13\%) & 95 (40\%) \\
\hline
\end{tabular}
\label{tbl:tdd}
\end{table}

\begin{table}
\caption{Implementation summary for FDD designs. In all the designs
  below, the input bandwidth is $800$~MHz.}
\begin{tabular}{l c c c c}
\hline
Parameter  & Design 1 & Design 2 & Design 3 & Design 4  \\
\hline
Mixer & No & No & Yes & Yes \\
Decimation factor & 8 & 10 & 8 & 10 \\
$B_w$ (MHz) & 100 & 80 & 100 & 80 \\
$C_f$ (MHz) & 50 & 40 & 400 & 400 \\
Latency (ns) & 35 & 440 & 50 & 455 \\
FPGA slices (Out of 23616) & 4175 (17\%) & 6142 (26\%) & 5690 (24\%) & 6439 (27\%)\\
4 input LUTs (Out of 47232) & 5153 (10\%) & 5216 (11\%) & 5984 (12\%)  & 6003 (12\%) \\
Block RAMs (Out of 232) & 41 (17\%) & 41 (17\%) & 49 (21\%) & 49 (21\%)\\
$18 \times 18$ Multipliers (Out of 232) & 8 (3\%) & 8 (3\%) & 32 (13\%) & 32 (13\%)\\
\hline
\end{tabular}
\label{tbl:trade-off}
\end{table}

\begin{table}
\caption{Implementation summary for designs employing two-stage
  downconversion using cascaded FDF or TDF blocks. In all the designs
  below, the input bandwidth is $800$~MHz. None of these designs
  employs a mixer block.}
\begin{tabular}{l c c c c}
\hline
Parameter  & Design 1 & Design 2 & Design 3 & Design 4  \\
\hline
No. of FDF blocks & 0 & 2 & 1 & 1 \\
No. of TDF blocks & 2 & 0 & 1 & 1 \\
FDF Decimation factor(s) & --- & 8, 10 & 8 & 10 \\
$B_w$ (MHz)$^{**}$ & Tunable & 10 & Tunable &  Tunable \\
& ($\leq 800$) &  & ($\leq 100$) & ($\leq 80$) \\
Latency (ns) & 170 & 475 & 120 & 505 \\
FPGA slices (Out of 23616) & 17141 (72\%) & 5765 (24\%) & 11073 (46\%) & 12641 (53\%)\\
4 input LUTs (Out of 47232) & 19718 (41\%) & 5506 (11\%) & 12245 (25\%)  & 12310 (26\%) \\
Block RAMs (Out of 232) & 41 (17\%) & 41 (17\%) & 41 (17\%) & 41 (17\%)\\
$18 \times 18$ Multipliers (Out of 232) & 64 (27\%) & 16 (6\%) & 40 (17\%) & 40 (17\%)\\
\hline
\end{tabular}
\tablenotetext{**}{$B_w$, if tunable, can be tuned to frequencies
  consistent with decimation factors supported by the TDD block.}
\label{tbl:cascading}
\end{table}

\end{document}